# The non-linear trade-off between return and risk: a regime-switching multi-factor framework

John Cotter [*]

*Financial Mathematics and Computation Cluster (FMC2), Smurfit School of Business, University College Dublin, Blackrock, Co. Dublin, Ireland*
*Telephone number: +353 1 7168900/ e-mail: john.cotter@ucd.ie;*

Enrique Salvador [**]

*Financial Mathematics and Computation Cluster (FMC2), Smurfit School of Business, University College Dublin, Blackrock, Co. Dublin, Ireland*
*Telephone number: +353 1 7164378/ e-mail: enrique.salvador@ucd.ie;*
*and Universitat Jaume I, Finance and Accounting Department*
*Avda. Sos Bynat s/n, E-12071 Castellon de la Plana, Spain*

**Abstract**

This study develops a multi-factor framework where not only market risk is considered but also potential changes in the investment opportunity set. Although previous studies find no clear evidence about a positive and significant relation between return and risk, favourable evidence can be obtained if a non-linear relation is pursued. The positive and significant risk-return trade-off is essentially observed during low volatility periods. However, this relationship is not obtained during periods of high volatility. Also, different patterns for the risk premium dynamics in low and high volatility periods are obtained both in prices of risk and market risk dynamics.

[**]Corresponding author

# 1. Introduction

The relation between expected return and risk has motivated many studies in the financial literature. Most recent asset pricing models are based on this fundamental trade-off, so understanding the dynamics of this relation is a key issue in finance. One of the first studies establishing a theoretical relation between expected return and risk is the Sharpe (1964) and Lintner (1965) CAPM. These authors propose a positive linear relationship between the expected return of any asset and its covariance with the market portfolio. Later, Merton (1973) proposed an extension of this model, adding a second risk factor in the relationship that may improve the static CAPM. The market risk premium in the ICAPM is proportional to its conditional variance and the conditional covariance with the investment opportunity set (hedging component).

The empirical literature on the ICAPM has tested the implications of this model in two dimensions. Papers such as Shanken (1990), Brennan et al. (2004) Ang et al. (2006), Lo and Wang (2006), and Petkova (2008) focus on the cross-sections of excess stock returns. Most of these works include additional factors other than market returns to provide an improved description of the dispersion in excess portfolio returns in the cross-section. Other related studies on the ICAPM with a focus on the time-series aggregate risk-return trade-off can be found in papers such as Scruggs (1998), Whitelaw (2000), Brandt and Kang (2004), Ghysels et al. (2005) and Guo et al (2009). These papers try to unmask a fundamental relationship between return and risk in financial data using different time-series analysis techniques.

This study aims to shed light on the empirical validation of the risk-return trade-off following the time-series dimension of the second set of studies. There are many studies analyzing this relationship empirically, but their results are controversial. Campbell (1987), Glosten et al. (1993), Whitelaw (1994) and Brandt and Kang (2004) find a negative relation between these variables[1], while other authors, such as Ghysels et al. (2005), Leon et al. (2007), Guo and Whitelaw (2006), Ludvigson and Ng (2007) and Lundblad (2007), find a positive trade-off. There are others studies, such as Baillie and De Gennaro (1990) and Campbell and Hentschel (1992), that find non-significant estimates for this risk-return trade-off.

Several assumptions are necessary for the theoretical model to empirically analyze the aggregated risk-return trade-off in the time-series dimension. The most common is to consider constant prices of risk (Goyal and Santa-Clara 2003, Bali et al 2005). It is also necessary to assume specific dynamics for the sources of risk in the model. Finally, the empirical model is established in a discrete time economy instead of the continuous time economy used in the equilibrium model of the theoretical approach. Most of the empirical papers studying the risk-return trade-off use one or more of these assumptions.

Although a large body of literature focuses on this empirical validation, only a few studies use multi-factor models that consider a stochastic investment opportunity set (use of a hedge component). One of the most common simplifications when empirically analyzing this risk-return relationship is the consideration of a constant set of investment opportunities (Glosten et al 1993, Lundblad 2007), or alternatively, independent and identically distributed returns. This assumption implies that the market risk premium only depends on its conditional variance and could be validated using single rather than multi-factor models.

The great controversy in the empirical validation of the risk-return trade-off is due to the disappointing results obtained about the sign and significance of this relation. There is no consensus about whether these results are due to: (1) wrong specifications of conditional volatility and the dynamics of risk factors (Guo and Neely (2008), Leon et al. (2007)); (2) misspecifications of the empirical models caused by the omission of the hedge component (Scruggs (1998)); (3) or both.

---

[1] See Abel (1988) and Backus and Gregory (1992) for theoretical models that support a negative risk-return relation.

The empirical technique most used in the literature to show this trade-off is the GARCH-M framework. This methodology assumes a linear relation between return and risk. There are other approaches to empirically analyzing the risk-return trade-off. However, most of them use different econometric techniques to validate a linear relationship between return and risk based on Merton's ICAPM model. For instance, Ghysels et al. (2005) use the MIDAS regression, Ludvigson and Ng (2007) use factor analysis with macroeconomic variables and Bali and Engle (2010) use a temporal and cross-sectional analysis for a wide range of portfolios comprising the whole market.

In this paper, we use another econometric approach based on the equilibrium model of Whitelaw (2000) in which we consider a non-linear relationship between return and risk. Generally, theoretical models do no restrict the risk-return relation to be linear or monotonic (Rossi and Timmerman (2010), Campbell and Cochrane (1999), Whitelaw (1994), Harvey (2001)). It is shown in this paper that the relationship between the expected return and volatility follows a non-linear rather than linear pattern, as stated in the ICAPM model.

We use a Regime Switching GARCH (RS-GARCH) approach that allows us obtain favorable evidence for a positive and significant risk–return trade-off. We present a multi-factor model (assuming a stochastic set of investment opportunities) where both the prices and sources of risk are state-dependent. We consider non-linear relationships between return and risk following the papers of Mayfield (2004) and Whitelaw (2000).

Using monthly excess market return data from 1953 to 2013, we find a positive and statistically significant relationship between return and risk during low-volatility states. The evidence for this relationship during high-volatility states is less clear and is not significant when we assume a linear risk-return trade-off. This result may give an answer to the risk-return puzzle. Given the evidence in this paper, we show that the consideration of different states in the market allows us to unmask this fundamental trade-off. The results show that there are periods in low-volatility states when a positive and significant relationship between return and risk exists, supporting the theoretical model. However, there are other periods in high-volatility states when this trade-off is not observed or is even negative. Thus, the assumption of a linear risk-return trade-off may fail to uncover this fundamental relationship because this trade-off depends on the state of the market.[2]

The results in this paper also present differences in the patterns followed by both prices of risk and conditional volatilities in the different states. One interesting result is that the magnitude of the market price of risk is lower during high-volatility states. These results suggest a pro-cyclical behavior in the investor risk-appetite, which depends on the volatility regime (high/low volatility). During low-volatility states, investors ask for a higher price for a 'unit' of risk than during high-volatility states. Additionally, the dynamics of conditional volatilities show greater persistence during high-volatility regimes with non-stationary dynamics during periods of financial turmoil.

---

[2] To illustrate, the use of a timeframe with many periods corresponding to low-volatility states would result in a positive and significant risk-return trade-off. However, if, in the chosen timeframe, there were a high number of high-volatility states, they would result in a negative and significant trade-off. Non-significant estimations for the relation between return and risk would likely be found in samples with a similar number of high and low-volatility periods.

Although the previous results seem to go against the spirit of theoretical models, the same results are obtained in several papers using alternative methodologies. Evidence for lower prices of risk during high-volatility periods can be found in Bliss and Panigirtzoglou (2004). These authors use a utility function to adjust the risk-neutral probability distribution function embedded in options. They are able to obtain measures of the risk aversion implied in options, and they find that the degree of relative risk aversion is lower during periods of high volatility. Regarding the state-dependent relationship between return and risk, Rossi and Timmerman (2010) use a flexible econometric approach that shows a non-monotonic relation between conditional volatility and expected market returns (at low-medium levels of volatility, a positive risk-return trade-off is observed, but this relation becomes inverted at high levels of volatility).

Our goal in this paper is not to question the use of theoretical models in representing empirical features of the market. In fact, we want to highlight that the ICAPM is still alive and that the fundamental relationship between return and risk can be observed in the markets. However, this is only true for certain periods that correspond with low-volatility states. In contrast, in periods of instability and financial turmoil, the observed trade-off between return and risk follows other patterns.

In this paper, the risk-return relation is tested using different proxies for the market portfolio (including the CRSP Value-Weighted NYSE-AMEX index and the SP500 index) and using datasets similar to those of previous studies (Scruggs, 1998 and Scruggs and Glabadanidis, 2003). The robust evidence obtained in all cases supports our main conclusions and highlights the potential perils of making linear assumptions when modeling state-dependent markets.

The rest of the paper is organized as follows. Section 2 provides a description of the data. Section 3 develops the empirical framework used in the paper. Section 4 presents the main empirical results, followed by robustness results in Section 5. Section 6 analyses the evidence after controlling for high volatility observations, and Section 7 concludes.

## 2. Data Description

This study uses 724 monthly excess market returns from the US market, including observations from March 1953 to June 2013. The application of our models is constructed around the availability of data on Government Bonds from the Federal Reserve Bank of St. Louis, which starts in 1953. This dataset is large and similar in size to other studies such as Scruggs (1998) or Scruggs and Glabadanidis (2003). Although there are slight differences in the parameter estimates using different data frequencies (monthly, weekly or daily), there is no particular reason why the conclusions in this study should be affected by the choice of data frequency (see, e.g., De Santis and Imhoroglu 1997).

We compute excess market returns as the difference in the total returns of the CRSP Value-Weighted NYSE-AMEX index minus the risk-free rate. Following Scruggs (1998) and Mayfield (2004), the yield of the 1-month T-bill is used as the proxy for the risk-free rate. The chosen proxy for the hedging component against changes in the investment opportunity set is the total returns for a set of constant-maturity US Government Bonds (similar to Bali and Engle (2009, 2010)): 5-, 10-, and 20-year Treasury bonds and an equally averaged portfolio containing these three bonds. The Centre for Research in Security Prices (CRSP) database is used to obtain the market portfolio return data. The Federal Reserve Bank of St. Louis provides the data corresponding to the yields on the risk-free rate and the yields on the proxies used as the intertemporal hedging component. To compute the total return on a constant maturity bond from the bond yield, we add two components: the promised coupon at the start of the year and the price change due to interest rate changes (see Morningstar® Bond Return Calculation Methodology (2013) for details).

Figure 1 plots the returns for the excess market returns, the risk-free rate and the excess returns for the alternative investment possibilities. Table 1 shows the main summary statistics for excess returns in the market portfolio and the intertemporal hedging alternatives and their correlations.

[INSERT FIGURE 1]

[INSERT TABLE 1]

The results show that any of the selected series could be considered a proxy reflecting the alternative investment set available to investors. We observe a very small correlation between any of the hedging component variables and the excess market returns. Nevertheless, due to the lack of consensus in the literature about the best proxy representing the alternative investment set (Scruggs and Glabadanidis 2003, Guo and Whitelaw (2006), Bali 2008), this study uses all of these different assets presenting different characteristics (in their terms and maturities), allowing us to bolster the robustness of our study.

The excess monthly market returns have a mean of 0.567% and a standard deviation of 4.21%. Excess market returns present evidence of negative skewness and leptokurtosis. These results, together with the Jarque-Bera test, suggest that the excess market returns follow non-normal distributions with evidence of fat tails. The excess bond returns for the different alternatives chosen vary from 0.123% in the 5-year Government Bond to 0.172% in the 20-year Government bond and a standard deviation of 1.271% in the 5-year bond to 2.415% in the 20-year bond (the longer the bond maturity, the larger the mean and standard deviation).

Furthermore, all series exhibit conditional heteroskedasticity (serial autocorrelation in squared returns), and there is evidence of volatility clustering from the time-series plots. With these serial correlation patterns, the use of GARCH models to represent the dynamics of conditional second moments, which has large support in the previous literature, is understandable.[3]

## 3. Empirical Methodology

We now present the empirical models used in the study. One of the main building blocks of this paper is the assumption of state-dependent prices of risk and state-dependent conditional volatilities. This modeling implies a non-linear relationship between return and risk following the equilibrium model in Whitelaw (2000). We assume bivariate GARCH dynamics for conditional volatilities, more specifically, the BEKK model of Baba et al (1990). The main advantage of this model is that it guarantees that the covariance matrix will be positive definite by construction (quadratic form). State-independent multi-factor models that establish a linear relation between return and risk are presented in Section 3.1. In Section 3.2, state-dependent multi-factor models are introduced. These models establish a non-linear risk-return trade-off through a regime-switching process both in the risk premium and conditional volatilities.

---

[3] Although there is also some serial autocorrelation in the series levels, we do not consider the inclusion of any structure in the mean equation because GARCH modeling eliminates any serial correlation in the standardized residuals. Standardized residual results are available on request.

## 3.1. State-independent multi-factor model

We present a multi-factor model derived from Merton's (1973) ICAPM model. The market risk premium in the theoretical Merton model is defined as:

$$E_t(R_{W,t}) = \left[\frac{-J_{WW}W}{J_W}\right]\sigma_{W,t}^2 + \left[\frac{-J_{WB}}{J_W}\right]\sigma_{WB,t} \quad (1)$$

where $J$ is the utility function (subscripts representing partial derivatives), $W$ is the wealth level, $B$ is a variable that describes the state of investment opportunities in the economy, $E_t(R_{W,t})$ is the expected excess return on aggregate wealth, $\sigma_{W,t}^2$ and $\sigma_{WB,t}$ are, respectively, the conditional variance and the conditional covariance of the excess returns with the investment opportunity set, and $\left[\frac{J_{WW}W}{J_W}\right], \left[\frac{J_{WB}}{J_W}\right]$ are the prices of the risk factors.

Using this model, we empirically test the aggregated (linear) risk-return trade-off in the time-series dimension. The 'general' model allows time-varying conditional second moments, but the price of risk coefficients for market risk $\left[\frac{J_{WW}W}{J_W}\right]$ and intertemporal component risk $\left[\frac{J_{WB}}{J_W}\right]$ are constant over time (Scruggs and Glabadanidis 2003). Thus, given the theoretical framework and the assumptions taken, the empirical model relating excess market returns with the risk factors is stated as follows:

$$\begin{aligned} r_{m,t} &= \lambda_{10} + \lambda_{11}\sigma_{m,t}^2 + \lambda_{12}\sigma_{mb,t} + \varepsilon_{m,t} \\ r_{b,t} &= \lambda_{20} + \lambda_{21}\sigma_{mb,t} + \lambda_{22}\sigma_{b,t}^2 + \varepsilon_{b,t} \end{aligned} \quad (2)$$

where $r_{m,t}$ and $r_{b,t}$ are the excess market returns on the market portfolio and the alternative investment set, respectively; $\lambda_{ij}$ for $i=1,2$ and $j=0,1,2$ are the parameters to be estimated, which represent the different prices of risk; and $\sigma_{m,t}^2$, $\sigma_{b,t}^2$, $\sigma_{mb,t}$ represent the conditional second moments (market variance, intertemporal hedging component variance and covariance between market portfolio and hedging component) or risk factors. A 'restricted' version of this model is also estimated, where the alternative investment set is time invariant ($\lambda_{21} = \lambda_{22} = 0$) (similar to Scruggs 1998).

As explained above, it is necessary to make an assumption about the dynamics of the volatilities (risk factors) in order to empirically validate the theoretical ICAPM model. To analyze bivariate relationships, one of the most widely used models in the literature is the BEKK model of Baba et al. (1990). This model sets the following covariance equation:

$$H_t = \begin{pmatrix} \sigma_{m,t}^2 & \sigma_{mb,t} \\ \sigma_{mb,t} & \sigma_{b,t}^2 \end{pmatrix} = CC' + A'\varepsilon_{t-1}\varepsilon_{t-1}'A + B'H_{t-1}B \quad (3)$$

where $C$ is a lower triangular $2x2$ matrix of constants, $A$ and $B$ are $2x2$ diagonal matrices of parameters[4], $\varepsilon_{t-1}$ is a $Tx2$ vector of innovations and $H_{t-1}$ is the lagged covariance matrix.

---

[4] Diagonal BEKK models are more parsimonious than full models and they perform well in representing the dynamics of variances and covariances (Bauwens et al 2006).

The model is estimated by the maximization of the Quasi-Maximum Likelihood function, assuming that the innovations follow a normal bivariate distribution: $\varepsilon_t \sim N(0, H_t)$.

$$L(\theta) = \sum_{t=1}^{T} \ln\left[f(r_t, \Omega_t; \theta)\right] \quad \text{where} \quad f(r_t, \Omega_t; \theta) = (2\pi)^{-1}|H_t|^{-\frac{1}{2}}\exp\left(-\frac{1}{2}\varepsilon_t' H_t^{-1}\varepsilon_t\right) \quad (4)$$

where $|H_t|$ represents the determinant of the covariance matrix, $\Omega_t$ is the information set up to $t$ and $\theta$ is the vector of unknown parameters.

## 3.2. Regime-switching multi-factor model

We now introduce a different multi-factor model where both the prices of risk and the conditional second moments are dependent on the state of the market. In this case, we propose two states[5]. The consideration of regime-switching in the empirical relation allows us obtain state-dependent estimations for the risk prices and conditional second moments. This implies a non-linear and state-dependent relation between expected return and risk following the general equilibrium model developed in Whitelaw (2000). Here, the study on the existence of a fundamental trade-off between return and risk is conditioned within each regime (defined by the levels of volatility in the market). This modeling gives us the chance to analyze a different and more complex shape of the risk-return relation beyond the simple linear one stated in the theoretical ICAPM model.

With these assumptions, the mean equation specification in this model is defined as:

$$\begin{aligned} r_{m,t,s_t} &= \lambda_{10,s_t} + \lambda_{11,s_t}\sigma^2_{m,t,s_t} + \lambda_{12,s_t}\sigma_{mb,t,s_t} + \varepsilon_{m,t,s_t} \\ r_{b,t,s_t} &= \lambda_{20,s_t} + \lambda_{21,s_t}\sigma_{mb,t,s_t} + \lambda_{22,s_t}\sigma^2_{b,t,s_t} + \varepsilon_{b,t,s_t} \end{aligned} \quad (5)$$

where $r_{m,t,st}$ and $r_{b,t,st}$ are state-dependent (on the latent variable $s_t=1,2$) excess market and hedging component returns, respectively; $\lambda_{ij,s_t}$ for $i=1,2$ and $j=0,1,2$ are state-dependent parameters; $\sigma^2_{m,t,s_t}$, $\sigma^2_{b,t,s_t}$ and $\sigma_{mb,t,s_t}$ are the state-dependent conditional second moments; and $\varepsilon_{m,t,s_t}$ and $\varepsilon_{b,t,s_t}$ are the state-dependent innovations.

It is assumed that the state-dependent conditional second moments follow GARCH bivariate dynamics (more specifically, a BEKK model). That is, there are as many covariance matrices as states.

The state-dependent covariance matrices are:

$$H_{t,s_t} = \begin{pmatrix} \sigma^2_{m,t,s_t} & \sigma_{mb,t,s_t} \\ \sigma_{mb,t,s_t} & \sigma^2_{b,t,s_t} \end{pmatrix} = C_{s_t} C_{s_t}' + A_{s_t}' \varepsilon_{t-1} \varepsilon_{t-1}' A_{s_t} + B_{s_t}' H_{t-1} B_{s_t} \quad (6)$$

where (for $s_t=1,2$) $A_{st}$ and $B_{st}$ are 2x2 diagonal matrices of parameters, and $C_{st}$ are 2x2 lower triangular matrices of constants.

---

[5] Previous studies considering three states (e.g., Sarno and Valente 2000) when modeling stock and futures time-series show that the third state only reflects odd jumps in the return series. The explanatory power of this third state has been found to be low.

Shifts from one regime to another are governed by a hidden variable that follows a first-order Markov process with probability transition matrix $P$ (Hamilton 1989).

$$P = \begin{pmatrix} \Pr(s_t = 1 | s_{t-1} = 1) = p & \Pr(s_t = 1 | s_{t-1} = 2) = (1-q) \\ \Pr(s_t = 2 | s_{t-1} = 1) = (1-p) & \Pr(s_t = 2 | s_{t-1} = 2) = q \end{pmatrix} \quad (7)$$

where $p$ and $q$ are the probabilities of being in states 1 and 2 if in the previous period the process was in states 1 and 2, respectively.

Due to this state-dependence and the recursive nature of GARCH models, the construction and estimation of the maximum likelihood function would be intractable unless independent estimates for innovations and covariances were obtained. To solve this problem, we use a recombinative method similar to that used in Gray (1996). This method allows us obtain state-independent estimations for the covariance matrix and the innovations by weighting the state-dependent covariance matrix and innovations with the ex-ante probability of being in each state.

The ex-ante probabilities (the probabilities of being in each state in period $t$ using the information set at $t-1$) are given by (8) and (9):

$$P(s_t = 1 | \Omega_{t-1}; \theta) = p * P(s_{t-1} = 1 | \Omega_{t-1}; \theta) + (1-q) P(s_{t-1} = 2 | \Omega_{t-1}; \theta) \quad (8)$$

$$P(s_t = 2 | \Omega_{t-1}; \theta) = 1 - P(s_t = 1 | \Omega_{t-1}; \theta), \quad (9)$$

where

$$P(s_t = k | \Omega_t; \theta) = \frac{P(s_t = k | \Omega_{t-1}; \theta) f(r_t | s_t = k, \Omega_t; \theta)}{\sum_{k=1}^{2} P(s_t = k | \Omega_{t-1}; \theta) f(r_t | s_t = k, \Omega_t; \theta)} \quad (10)$$

and $k=1, 2$ are the filtered probabilities (the probabilities of being in each state in period $t$ with the information set up to $t$).

Assuming state-dependent innovations following a normal bivariate distribution $\varepsilon_{t,s_t} \sim N(0, H_{t,s_t})$, the vector of unknown parameters $\theta$ is estimated by maximizing the following maximum-likelihood function:

$$L(\theta) = \sum_{t=1}^{T} \ln \left[ \sum_{k=1}^{2} P(s_t = k | \Omega_t; \theta) f(r_t, \Omega_t; \theta) \right] \quad \text{where} \quad f(r_t, \Omega_t; \theta) = (2\pi)^{-1} |H_t|^{-\frac{1}{2}} \exp\left(-\frac{1}{2} \varepsilon_t' H_t^{-1} \varepsilon_t\right) \quad (11)$$

where the state-dependent likelihood function is weighted by the ex-ante probability of being in each state.

In this state-dependent model, we cannot make a direct interpretation of the magnitude of the risk-aversion coefficient for the representative investor. Unlike the single-regime multi-factor model presented here, the coefficient accompanying the market risk factor cannot be viewed as investor risk aversion but as the market price of risk (see Merton 1973, Whitelaw, 2000). However, we can make approximations for the risk aversion in terms of the sign of this price of risk and the relative comparisons of this price across states.

Implementing these two types of models, linear and non-linear, allows us re-examine the patterns followed by the risk-return relation in the US market and to shed light on the empirical controversy about its sign and significance. Detailed estimates and discussions of these results are provided in the next section.

## 4.- Empirical Results

We turn to the empirical results for the models proposed in the previous section. We estimate these models with the different proxies used for the intertemporal hedging component. Model I (both linear and state-dependent) use the 5-year T-bond as a proxy for the intertemporal hedging component. A similar approach is taken for model II with the 10-year T-bond, model III with the 20-year T-bond and model IV with the equally weighted bond portfolio.

Section 4.1 shows and discusses the results for the linear models (without regime-switching), both in the general and restricted cases. Section 4.2 explains the results for the non-linear multi-factor models (general and restricted), including regime switching, and their implications for trying to unmask the aggregated risk-return trade-off in the markets.

### *4.1.- Multi-factor model estimations*

The estimated models are those introduced in section 3.1. We estimate a restricted version of the models where we assume constant risk premiums for the hedge component, i.e., $\lambda_{21} = \lambda_{22} = 0$. In the general version of the model, we estimate all parameters freely. The estimated parameters for the mean equation are presented in Table 2.A.

**[INSERT TABLE 2]**

Most of the parameters in the mean equation for this multi-factor model are non-significant. The coefficients that reflect the market price of risk ($\lambda_{11}$) are positive but non-significant in all cases considered. Similar results are obtained for the hedging component risk factor ($\lambda_{12}$). This means our model fails to detect a significant relation between return and risk if it is assumed to be linear in the whole sample. This result is similar to other studies for the risk-return relation (Baillie and di Gennaro. 1990; Campbell and Hentschel, 1992), which also find a non-significant relationship when using similar linear models.

Table 2.B shows the parameter estimates for the variance equation. These parameters define the dynamics and patterns followed by the conditional second moments. The bivariate GARCH specification is a good fit and properly captures the conditional second moment dynamics[6]. Significance in the parameters representing shocks in volatility ($a_{11}$, $a_{22}$) and persistence of past variance ($b_{11}$, $b_{22}$) is observed for both risk factors (market risk and the investment opportunity set component). The persistence level in the two sources of risk—market risk ($b_{11}$) and hedging component ($b_{22}$)—is relatively high using multi-factor models, with values close to 1. This high persistence level suggests the presence of several regimes in the volatility process, in agreement with Lameroux and Lastrapes (1990). These authors state that persistence in variance may be overstated because of the existence, and failure to take account of, regime changes in the model.

Thus, ignoring these regime shifts could lead to inefficiency in the volatility estimates and, consequently, in the risk factors. Regime-Switching (RS)-GARCH models allow us to both consider different states of the volatility process, as we explain in the next sub-section, and to overcome this limitation. More importantly, this type of model allows us to analyze more complex shapes of the empirical risk-return trade-off that cannot be detected by simpler linear models.

---

[6] Diagnostics tests are not reported for brevity but are available on request.

*4.2- Regime-Switching multi-factor model estimation*

We show estimates for the state-dependent models presented in Section 3.2. These models exhibit state-dependent prices of risk and conditional second moments. Table 3 describes the estimates for the state-dependent mean equation in all cases considered. As we explain below in Figure 3, we can associate states 1 and 2 with low and high-volatility periods, respectively.

**[INSERT TABLE 3]**

Panel A of Table 3 shows the mean equation results for low-volatility periods (state 1). Positive and significant estimates for the market price of risk in low-volatility states ($\lambda_{11,s=1}$) are obtained in all cases considered (for all proxies used as the intertemporal hedging component both in the general and restricted versions of the model). However, the magnitude of the estimated coefficients of these prices of risk depends on the model. Additionally, the results for the intercept are non-significant in most cases, following theoretical intuition. The original equilibrium framework does not include an intercept term, but empirical models include it to test for potential model misspecification. The non-significance of this term provides more support for the equilibrium model in these low-volatility states. A positive and significant influence of the covariance between the risk premium and the hedging component ($\lambda_{12,s=1}$) on the market risk premium is also observed in some specifications. The evolution of this covariance does not exhibit a clear positive or negative influence on the total risk premium demanded (see Figure 2). The estimated product of the price of risk times the covariance between excess market return and the hedging component $\left(\lambda_{12}\sigma_{mb,t,s_t=1}\right)$ implies that the total risk premium required by the investor $\left(\lambda_{11}\sigma^2_{m,t,s_t=1} + \lambda_{12}\sigma_{mb,t,s_t=1}\right)$ will be slightly lower than the market risk premium when this covariance is negative. However, when the covariance is positive, the premium associated with the hedging component will lead to higher values of the total risk premium.

**[INSERT FIGURE 2]**

Panel B of Table 3 shows the mean equation results obtained for state 2. There is no clear pattern for this regime. There is no significant relation between the expected return and risk in high-volatility states ($\lambda_{11,s=2}$) for most of the cases. However, in other cases, there is a negative risk-return trade-off at extremely high levels of volatility. This finding supports that of Rossi and Timmerman (2010). At low levels of conditional volatility, there is a positive trade-off between expected returns and risk, but this relationship becomes inverted at extremely high levels of volatility.

Moreover, the price of risk coefficients in state 1 (corresponding to low-volatility states) are higher than those corresponding to state 2 (high-volatility states). Unlike the linear multi-factor models in section 4.1, we cannot directly associate the coefficients in this model with risk-aversion coefficients (see Merton 1973). However, we can define the willingness of investors to bear risk in a given environment. This concept depends upon both the degree to which investors dislike uncertainty and the level of uncertainty. In this way, we can illustrate the difference between the risks perceived by the agents at a given point in time by risk aversion itself. Our estimated price of risk measures how much an investor would pay for one 'unit of risk' under a certain environment. Thus, if the price of risk is taken together with the quantity of risk inherent in the market, one is calculating the expected return to compensate investors for trading in the market, the well-known risk premium (we discuss this issue further in section 5.2).

Although we cannot use the parameters as risk aversion coefficients in our state-dependent multi-factor models, we can establish certain relationships with our findings. Prices of risk reflect the amount an investor would pay for a 'unit of additional risk'. Intuition tells us that a risk-averse investor would ask for a higher price of risk for the same amount of risk than a risk-loving investor. Therefore, we can draw a direct positive relation between the price of risk and risk aversion in our specification. If, then, we obtain higher price of risk estimates in our model, we can associate it with increases in risk aversion levels in the market. On the other hand, when we obtain lower price of risk, we can associate it with decreases in risk aversion levels in the market.

The previous discussion suggests that there is a lower risk aversion level in high-volatility states. This finding is inconsistent with the spirit of linear theoretical models, which suggests that higher volatility should be compensated with higher returns. One potential explanation for this result could be the different risk aversion profiles for investors in each state (Bliss and Panigirtzoglou, 2004). Using options on the S&P100 and the S&P500, Bliss and Panigirtzoglou (2004) find that risk aversion is higher during periods of low volatility. These authors do not give an explanation for this result; they simply recommend developing theoretical models to capture these effects. Given our results, we would concur that during calm (low-volatility) periods, more risk-averse investors trade in markets. However, in high-volatility periods, only the less risk-averse investors remain in the market because they are the only investors interested in assuming such risk levels, thereby decreasing the price of risk demanded during these periods. This interpretation is in line with recent papers such as Salvador at el. (2014) and Ghysels et al (2013). These papers document that the Merton model holds over samples excluding financial crises, considering these periods as "flight-to-quality" regimes. The separation of the traditional risk-return relation from financial crises leads to fundamental changes in the relation.

Furthermore, other related papers, such as Kim and Lee (2008), have reported similar evidence to ours in obtaining a significant risk-return trade-off during boom periods that is less clear during crisis periods. In this study, we do not define the states of the economy depending on the business cycle (boom/crisis), but we use volatility regimes. However, one would expect a link between volatility regimes and business cycles, which we will discuss in more detail later. Low-volatility states would correspond to boom periods, while the less-common high-volatility states would be associated with crisis periods (Lundblad, 2007). The procyclical risk-aversion (investors show more risk-aversion during boom periods than during crisis periods) documented in Kim and Lee (2008) is replicated in our approach using volatility regimes, where investors show more risk-aversion during low-volatility periods than during high-volatility periods.

Table 4 shows the estimations for the state-dependent variance equations. Again, significant estimates are obtained for the parameters accompanying the shock impact ($a_{11}$, $a_{22}$) and persistence ($b_{11}$, $b_{22}$) in the volatility formation in both risk factors.

**[INSERT TABLE 4]**

Furthermore, volatility formation depends on the regime considered in this framework. For low-volatility regimes, a lower influence of the lagged variance (matrix $B_1$) is observed, even compared to the non-switching case (with values lower than unity in all cases). Moreover, in these states, there is usually a higher impact of shocks (matrix $A_1$) in volatility formation. In this case, the volatility observed at a period $t$ in a low-volatility state is determined less by the variance observed in the previous period than by the shock occurring in period $t$.

However, there is a decrease of the shock influence on volatility formation in high-volatility regimes ($a_{11,s_t=2}$, $a_{22,s_t=2}$). There is also an increase in the volatility persistence in these high-volatility states ($b_{11,s_t=2}$, $b_{22,s_t=2}$), with values higher than unity in most cases. Although volatility dynamics seem non-stationary during these highly volatile periods, the whole variance process remains stationary (Abramson and Cohen, 2007). What these results suggest is that the volatility process in the markets during periods of market turmoil could be explosive and may not necessarily revert to an equilibrium point. However, the whole process remains stationary because the volatility process will eventually return to the low-volatility state.

Thus, linear GARCH models could lead to overestimates of volatility persistence in low-volatility periods and underestimates of volatility persistence in high-volatility periods. This result can also affect the estimation of risk factors and, hence, the identification of a risk-return trade-off.

In addition, the non-linear multi-factor model allows us associate the different states that follow the volatility process with low- (*state 1*) and high-volatility (*state 2*) market periods. For instance, in the averaged portfolio case, the median of the estimated variances and covariances (scaled by $10^4$) for state 1 are $\hat{\sigma}^2_{m,s_t=1}$ = *8.5023*, $\hat{\sigma}^2_{b,s_t=1}$ = *1.6089* and $\hat{\sigma}_{mb,s_t=1}$ = *-0.0347*, while the medians of the series in state 2 are $\hat{\sigma}^2_{m,s_t=2}$ = *19.8916*, $\hat{\sigma}^2_{b,s_t=2}$ = *2.6144* and $\hat{\sigma}_{mb,s_t=2}$ = *-0.1215*. These results (jointly with Figure 3) allow us to associate the states defined in the non-linear model with low- (state 1) and high-volatility states (state 2).

Figure 3 shows the smooth probabilities[7] of being in state 1 (low volatility) during the sample period jointly with recessionary periods, as defined by the National Bureau of Economic Research (NBER).

### [INSERT FIGURE 3]

Although there are continuous changes in regime during the sample period, we can draw some inferences. Usually, recession periods documented by the NBER coincide with periods governed by high-volatility states, but this is not always the case. Thus, although we cannot establish a direct relationship between high volatility and business cycles, they appear to be positively related (Lustig and Venderhal, 2012). Despite these continuous changes in regime, low-volatility regimes are more prevalent during the sample period. The number of monthly periods where the volatility process is in a low-volatility state (probability of being in a low-volatility state is higher than 0.5) is between 520 and 560 (depending on the proxy for the hedging component), corresponding to 71%-77% of the total sample. Thus, for approximately three-quarters of the sample, we do observe a positive and significant relationship between return and risk, but this relation is not observed a quarter of the time. However, if we make no distinction between regimes and we consider a linear relation across the whole sample period, this one-quarter of the time the relation is not observed blurs the evidence for the whole sample, leading to a conclusion of a non-significant trade-off.

The results obtained regarding the significance of the risk-return trade-off in both types of multi-factor models lead us to believe that the lack of empirical evidence in previous studies could be due to the assumption of a linear risk-return trade-off. Non-linear assumptions lead to favorable evidence for the risk-return trade-off in low-volatility states, but we cannot obtain

---

[7] The smooth probability is defined as the probability of being in each state considering the entire information set.
$$P(s_t=1|\Omega_T;\theta) = P(s_t=1|\Omega_t;\theta)\left[p\frac{P(s_{t+1}=1|\Omega_T;\theta)}{P(s_{t+1}=1|\Omega_t;\theta)}\right] + \left[(1-p)\frac{P(s_{t+1}=2|\Omega_T;\theta)}{P(s_{t+1}=2|\Omega_t;\theta)}\right]$$

favorable evidence when a linear trade-off is assumed. The consideration of the intertemporal component in the risk-return relation is of second-order relevance. We also obtain a significant impact of the intertemporal component in the risk-return relation, similar to Whitelaw (2000), but what truly unmasks the aggregated risk-return trade-off is the distinction across regimes.

## 5.- Robustness tests

### 5.1.- Alternative market portfolios

Let us reassess our previous evidence by considering different alternatives to the market portfolio. In addition to the CRSP Value-Weighted NYSE-AMEX index, Bali and Engle (2010) use other popular US stock indices to reflect market returns. Given the sample period in this paper (March 1953-June 2013), we choose the SP500 index[8]. To obtain the excess market returns, we compute the monthly logarithmic returns of the index (obtained from Bloomberg), and we subtract the risk-free rate.

[INSERT TABLE 5]

Panel A of Table 5 shows the results for the mean equation assuming a linear relationship between return and risk, where the equally weighted bond portfolio is considered the alternative investment set[9]. Similar to the results for the CRSP portfolio, the parameter identifying the relationship between market returns and risk is non-significant in all cases. The influence of the covariance between market returns and the alternative investment set is also non-significant. These results support the previous findings, where linear models are unable to identify the theoretical positive relationship between returns and risk.

The results for the mean equation in the regime-switching case are displayed in panel B of Table 5. Some findings here are worth mentioning. Again, we find a positive and significant relationship between return and risk during low-volatility states, supporting the favorable evidence obtained earlier for low-volatility periods. The effect of the covariance between the market portfolio and alternative investments tends to be negative for the SP500 index. For high-volatility states, the relation between return and risk is also significant, but in the case of the SP500, the relation becomes negative, suggesting an inverted trade-off during these periods.

This analysis again shows the importance of distinguishing between regimes and relaxing the linear assumption when providing evidence for a risk-return trade-off in financial markets. Regardless of which market portfolio is chosen, the empirical models are unable to detect a significant linear relationship between these two variables, but this trade-off does exist when a less restrictive non-linear framework is adopted.

### 5.2.- Risk premium evolution

As a further analysis, this section describes the evolution of the risk premium demanded by investors in the US, distinguishing between which proportions of the risk premium correspond to each risk factor, namely, the market risk and the hedging component. We compute the premium associated with market risk as the product of the price of risk by idiosyncratic risk $\lambda_{11}\sigma^2_{m,t}$ for linear multi-factor models (and similarly for the hedging component premium). For the non-linear case, this risk premium is obtained using the state-dependent market risk premium weighted by the smooth probability of being in each state $P(s_t=1|\Omega_T;\theta)\lambda_{11,s_t=1}\sigma^2_{m,t,s_t=1} + P(s_t=2|\Omega_T;\theta)\lambda_{11,s_t=2}\sigma^2_{m,t,s_t=2}$ (and similarly for the hedging component premium). The total risk premium is computed as the sum of the two factor premiums.

---

[8] Other indexes are available and not used. For example, for the NYSE index, the sample period available is shorter (January 1966-June 2013) because it was first introduced in December 1965. Also, the price-weighting construction of the DJIA index results discourages our use of it as a proxy for the market portfolio.
[9] We also estimate the models using the other alternatives for the hedging component and find similar results. Results are available on request.

For brevity, we only show results corresponding to the equally weighted bond portfolio as the alternative investment case.[10] Figure 4.A describes the market risk premium for the different alternatives of the market portfolio.

**[INSERT FIGURE 4]**

The market risk premium series share similar patterns for the sample period for all portfolios considered. For instance, there is a common rise in the market risk premium that coincides with high-volatility periods. There are several peaks that also coincide with recessionary periods, such as December 1969-Novemeber 1970 and November 1973-March 1975. A peak is also observed following the Asian crisis in 1987 and the turbulent period between 1998 and 2002 that ended with the bursting of the dot-com bubble. Finally, there is a common increase in the premium during the last financial crisis (2007-2009).

The median of the monthly risk premium series shows that over the past 60 years, the risk premium in the US has remained at approximately 3.5% to 4% per annum,[11] depending on the model and portfolio used. This estimate is very similar to other studies on US data (e.g., Bali (2008)). Furthermore, the total risk premium is essentially composed of the risk associated with the market. The percentage of the total risk premium corresponding to the hedging component is relatively small for both models.

To detect differences in the risk premium between the two proposed model types (linear and non-linear), Figure 4.B plots its evolution in each portfolio for both models[12]. A similar evolution of the total risk premium is observed in both the linear and non-linear models. It appears that during peaks of volatility, the non-linear models provide higher estimates of the premium; the rest of the time, they seem to provide lower estimates. These results suggest that during low-volatility periods, linear models tend to overestimate the premium, and during high-volatility periods they tend to underestimate it.

*5.3.- Datasets used in previous papers*

This section repeats our analysis for the datasets used in two well-known papers analyzing the risk-return trade-off. The papers of Scruggs (1998) and Scruggs and Glabadanidis (2003) represent two of the best-known references for the use of multi-factor models in trying to provide evidence in the time-series for Merton's theoretical model. Scruggs (1998) finds favorable evidence for a partial relation between the market risk premium and conditional market variance. He uses a bivariate model comprising the CRSP Value-Weighted NYSE-

---

[10] The dynamics of the evolution of the risk premium in the rest of the cases are very similar. Results are available on request.

[11] The descriptive statistics for the risk premiums are not shown but are available from the authors upon request.

[12] For brevity, only the figures for the average portfolio as an alternative investment in the general model are shown. The dynamics of the differences in the risk premium evolution in the rest of the cases are very similar. Results are available on request.

AMEX portfolio as the proxy for market returns ($r_{m,t}$) and long-term government bond returns (more specifically, the Ibboston Associates long-term U.S. Treasury bond total return index) as the hedging component factor ($r_{b,t}$) in a sample period from March 1950 to December 1994. A few years later, Scruggs and Glabadanidis (2003) tested whether the intertemporal variation in stock and bond premia can be explained by time-varying covariances with priced risk factors. Considering again the CRSP Value-Weighted NYSE-AMEX portfolio as market returns ($r_{m,t}$) and the Ibboston Associates long-term U.S. Treasury bond total return index as the second factor ($r_{b,t}$) in the models for January 1953 to December 1997, they use several bivariate models that relax the assumption of constant correlation in Scruggs (1998). However, they were unable to obtain favorable evidence for these types of models.

**[INSERT TABLE 6]**

Table 6 shows the results obtained for our empirical specifications using the same datasets as those in Scruggs (1998) and Scruggs and Glabadanidis (2003). If we analyze the results when using linear models (panel A), they are very similar to those of the datasets used in previous sections. We fail to identify a significant relation among these variables (even the covariances present no significance in any case).

Again, the evidence obtained using non-linear models is more favorable. We do obtain positive and significant estimations during low-volatility periods (panel B) in all cases as well as a robust negative effect of the covariance between market and bond returns for all cases. This implies that the market risk premia are positively related with the market variance but are reduced by the covariance between market and bond returns.

The results for high-volatility periods are also robust in all cases. Similar to some of the earlier specifications, it appears that the risk-return trade-off becomes inverted for high levels of volatility. In all cases, we obtain a highly negative and significant parameter. Additionally, going against the results for low-volatility states, the impact of the covariance in the market risk premium during periods of high volatility is positive. This implies that the market risk premia are negatively related to the market variance but are increased by the covariance between market and bond returns.

These results add further evidence regarding the perils of making linear assumptions when analyzing the risk-return trade-off in financial markets. The empirical evidence appears to support the theoretical model during low-volatility states, but the results do not hold for high levels of volatility.

## 6.- Do high-volatility regimes distort evidence on the risk-return trade-off?

Thus far, we have reported commonality in the results regardless of the dataset used or the model specification applied. We detect a significant positive risk-return trade-off under non-linear specifications during low-volatility states that we cannot find during high-volatility states or when we assume a simpler linear model. Thus, we suspect that during low-volatility periods, this trade-off does exist, but during high-volatility periods this relation is different, and aggregating these two different periods may distort the evidence in the entire sample. In this last section, we check whether we are able to detect a linear relationship between return and risk if we control for the observations associated with the high-volatility regimes.

We use different alternatives for the market portfolio (CRSP and SP500) and the equally weighted bond portfolio for the alternative investment set. We also use the datasets as in Scruggs (1998) and Scruggs and Glabadanidis (2003). From the original single-regime models in equation (2), we add two dummies to the equation of excess market returns (in levels and multiplying the market variance). These new dummies take the value 1 when the market is in periods of high volatility[13] and the value 0 the rest of the time. We discriminate between high and low volatility observations given the filtered probabilities estimated in sections 4.2 and 5.3.

The observations corresponding to a high-volatility state represent only a small percentage of the whole sample: only approximately 20% of the observations in the Scruggs (1998) and Scruggs and Glabadanidis (2003) datasets and even less (approximately 15% of observations) in the CRSP and SP500 portfolios.

Once we have the new dummies controlling for observations "within a high-volatility state," we run our proposed linear models described in Equations 2-4 (plus the two dummies) to test whether the existence of unusual observations corresponding to high-volatility states could have blurred previous evidence.

[INSERT TABLE 7]

Table 7 shows the results for this analysis. We can see that in all cases (CRSP, SP500, Scruggs (1998) and Scruggs and Glabadanidis (2003)), the non-significant estimations for the risk-return trade-off obtained previously using linear models turn out to be significant after controlling for the observations in high-volatility states. This suggests that the relationship between return and risk does exist during low-volatility states. However, the existence of periods of turmoil and financial instability may have masked this relationship in the time series dimension. Our results suggest a more complex relationship between risk and return than the linear relationship specified in the theoretical model. The negative and significant values for the dummy variables also reflect the inverted risk-return trade-off during high-volatility periods obtained in previous sections.

We also find non-significant estimates for the hedge component variable in the CRSP and SP500 indices when we control for high volatility observations. However, when we use datasets similar to those of Scruggs (1998) and Scruggs and Glabadanidis (2003), there is a negative and significant influence of the covariance on the market risk premium.

Thus, distinguishing among different patterns followed by risk premia and variances during periods of financial instability may help explain the puzzle of the risk-return trade-off. The evidence obtained supporting this claim in the US market is favorable for most cases. We suggest that this instability of the risk-return trade-off during high-volatility states has been a cause of the controversial results reported in previous studies.

## 7.- Conclusion

This paper empirically analyzes the risk-return trade-off for the US market under a multifactor framework using several types of US government bonds as alternative investment sets. We propose two different empirical models considering bivariate GARCH specifications that allow us identify linear and non-linear relationships between return and risk. The non-linear specification is obtained using a Regime-Switching model that allows us associate the different states with volatility regimes (low and high volatility).

Our main result shows that only a positive and significant risk-return trade-off is obtained in the non-linear case and only in the states governed by low volatility. However, we cannot find favorable evidence for the risk-return relation in the linear framework, and it is mixed during high-volatility states. Our results also support the findings of previous papers that document pro-cyclical behavior in the risk appetite of financial markets, i.e., during low-volatility states the market price of risk is higher than during high-volatility periods. The existence of periods where a positive and significant risk-return trade-off is not observed could lead to non-significant estimates for this relation for the entire sample. This result highlights the perils of using linear assumptions when analyzing the aggregated trade-off between return and risk and the inability of linear empirical models to capture a significant risk-return relationship, which exists most of the time.

---

[13] To be conservative, we only use high volatility observations that present a probability above 75%.

## Acknowledgements

The authors acknowledge the support of Science Foundation Ireland under Grant Number 08/SRC/FM1389 and the Irish Research Council under Grant Number GOIPD/2014/80. One of the authors also appreciates financial support from *Ministerio de Educación y Ciencia* project ECO2011-27227 and *UJI* project P1·1B2012-07. We appreciate comments from Michael Brennan, Thomas Conlon and Matthew Spiegel.

Table 1

Summary statistics for excess market returns and intertemporal hedging proxies

The table shows summary statistics and correlations of excess monthly returns for the market portfolio ($r_{m,t}$) and the alternative investment sets ($r_{b,t}$). The proxy for the market portfolio is the Center for Research in Security Prices (CRSP) value-weighted portfolio and the risk-free rate is the yield on the one-month Treasury bill. The proxies for the alternative opportunity set are the total returns (computed as Morningstar ®) for the 5, 10 and 20 year constant maturity US Government Bonds and an equally weighted portfolio using these 3 bonds. Panel A shows the mean, standard deviation, skewness, kurtosis for each series. It also shows the Jarque-Bera test for normality and the Ljung-Box test for serial autocorrelation in levels and squares using 6 lags *(t-stats in parenthesis)*. \*\*\*, \*\* and \* represent significance at 1%, 5% and 10% levels. The mean and standard deviation of each monthly series are expressed in percentage points. Panel B shows the correlations between the excess market returns and the different choices for the alternative investment set. The sample period includes observations from 1953 to 2013 and the returns are expressed as decimals.

| | Panel A.- Summary statistics | | | | |
|---|---|---|---|---|---|
| | ($r_{m,t}$) | ($r_{b,t}$) | | | |
| | Excess market return | Excess return 5-year T-bond | Excess return 10-year T-bond | Excess return 20-year T-bond | Excess return equally-weighted portfolio |
| Mean (x100) | 0.5672 | 0.1231 | 0.1495 | 0.1725 | 0.1492 |
| Std. Deviation (x100) | 4.2130 | 1.2713 | 1.8409 | 2.4154 | 1.5781 |
| Skewness | -0.4956 | 0.3615 | 0.4914 | 0.5904 | 0.6242 |
| Kurtosis | 5.0823 | 7.3289 | 6.8442 | 7.8485 | 6.9770 |
| J-B | 160.0143\*\*\* *(0.0000)* | 579.4733\*\*\* *(0.0000)* | 473.6377\*\*\* *(0.0000)* | 749.1486\*\*\* *(0.0000)* | 522.6930\*\*\* *(0.0000)* |
| L-B (6) | 26.5058 *(0.1492)* | 134.5243\*\*\* *(0.0000)* | 104.8266\*\*\* *(0.0000)* | 85.0971\*\*\* *(0.0000)* | 195.9049\*\*\* *(0.0000)* |
| L-B$^2$ (6) | 53.7536\*\*\* *(0.0000)* | 433.4219\*\*\* *(0.0000)* | 180.9236\*\*\* *(0.0000)* | 85.8342\*\*\* *(0.0000)* | 170.9475\*\*\* *(0.0000)* |
| Panel B.- Correlations | | | | | |
| | Excess market return | Excess return 5-year T-bond | Excess return 10-year T-bond | Excess return 20-year T-bond | Excess return equally-weighted portfolio |
| Excess market return | 1 | -0.0660 | -0.0606 | -0.0390 | -0.0013 |

Table 2.
Panel A Mean equation estimates for multi-factor models

The table shows the estimates of the risk-return trade-off using the independent multi-factor model in Equation 2. Two versions of the model are estimated: R refers to the Restricted version of the model in Equation 2 where $\lambda_{21} = \lambda_{22} = 0$ (Scruggs, 1998) and G to the General model where all parameters are freely estimated. Excess monthly returns are used in the estimation of the model. The proxy for the excess market returns ($r_{m,t}$) is the CRSP Value-Weighted portfolio minus the risk-free rate. Each column illustrates the results from the estimation of the model using one of the choices for the ($r_{b,t}$) alternative investment set. In model I the excess returns of the 5-year US Government Bond is used. In model II the excess returns of the 10-year US Government Bond is used. In model III the excess returns of the 20-year US Government Bond is used. In model IV the excess returns of an equally weighted portfolio containing these three bonds is used. The coefficients and corresponding t-statistics (in parenthesis) are shown for the full sample period (March 1953-June 2013). The t-statistics are computed using Bollerslev-Wooldridge standard errors (***, ** and * represents statistical significance at 1%, 5% and 10% levels).

| | | $r_{m,t} = \lambda_{10} + \lambda_{11}\sigma^2_{m,t} + \lambda_{12}\sigma_{mb,t} + \varepsilon_{m,t}$ | | | |
|---|---|---|---|---|---|
| | | $r_{b,t} = \lambda_{20} + \lambda_{21}\sigma_{bm,t} + \lambda_{22}\sigma^2_{b,t} + \varepsilon_{b,t}$ | | | |
| | | Model I | Model II | Model III | Model IV |
| $\lambda_{10}$ | R | -0.0034 (-0.5245) | -0.0083 (-0.8738) | -0.0074 (-0.8692) | -0.0068 (-0.8487) |
| | G | -0.0020 (-0.3255) | -0.0093 (-0.8325) | -0.0113 (-0.7580) | -0.0100 (-0.9397) |
| $\lambda_{11}$ | R | 5.8829 (1.5985) | 8.5568 (1.5533) | 7.7873 (1.5883) | 7.5177 (1.6205) |
| | G | 5.1010 (1.4604) | 9.1769 (1.4207) | 11.4738 (1.1258) | 9.4884 (1.5454) |
| $\lambda_{12}$ | R | -0.0992 (-0.0088) | -0.4122 (-0.0510) | 0.2648 (0.0447) | -0.1917 (-0.0238) |
| | G | -0.0819 (-0.0071) | -0.6355 (-0.0761) | 1.9528 (0.3155) | 0.7758 (0.0966) |
| $\lambda_{20}$ | R | 3.38e-04 (1.1448) | 3.38e-04 (0.9070) | 1.81e-04 (0.3565) | 2.28e-04 (0.6847) |
| | G | -2.19e-04 (-0.6094) | -3.54e-04 (-0.0778) | -2.06e-04 (-0.0227) | -6.71e-04 (-0.1632) |
| $\lambda_{21}$ | G | -1.3529 (-0.3375) | 1.7560 (0.3930) | 9.8629* (1.9512) | 7.5466* (1.7889) |
| $\lambda_{22}$ | G | 6.7558** (2.4735) | 3.9220** (2.1542) | 3.0509** (2.3168) | 5.4008*** (2.8771) |

.

Panel B Variance equation estimates for multi-factor models

The table shows the estimates of the variance equation of the single-regime multi-factor model in Equation 3. These variance equations come from two different versions of the main model presented in equation 2: R refers to the Restricted version of the model in Equation 2 where $\lambda_{21} = \lambda_{22} = 0$ (Scruggs, 1998) and G to the General model where all parameters are freely estimated. The proxy for the excess market returns ($r_{m,t}$) is the CRSP Value-Weighted portfolio minus the risk-free rate. Each column illustrates the results from the estimation of the model using one of the choices for the ($r_{b,t}$) alternative investment set. In model I the excess returns of the 5-year US Government Bond is used. In model II the excess returns of the 10-year US Government Bond is used. In model III the excess returns of the 20-year US Government Bond is used. In model IV the excess returns of an equally weighted portfolio containing these three bonds is used. Excess monthly returns are used in the estimation of the model. The coefficients and corresponding t-statistics (in parenthesis) are shown for the full sample period (March 1953-June 2013). The t-statistics are computed using Bollerslev-Wooldridge standard errors (***, ** and * represents statistical significance at 1%, 5% and 10% levels).

$$H_t = \begin{pmatrix} \sigma_{m,t}^2 & \sigma_{mb,t} \\ \sigma_{mb,t} & \sigma_{b,t}^2 \end{pmatrix} = \begin{pmatrix} c_{11} & 0 \\ c_{12} & c_{22} \end{pmatrix} \begin{pmatrix} c_{11} & 0 \\ c_{12} & c_{22} \end{pmatrix}' + \begin{pmatrix} a_{11} & a_{21} \\ a_{12} & a_{22} \end{pmatrix} \varepsilon_{t-1}\varepsilon_{t-1}' \begin{pmatrix} a_{11} & a_{21} \\ a_{12} & a_{22} \end{pmatrix}' + \begin{pmatrix} b_{11} & b_{21} \\ b_{12} & b_{22} \end{pmatrix} H_{t-1} \begin{pmatrix} b_{11} & b_{21} \\ b_{12} & b_{22} \end{pmatrix}'$$

| | | Model I | Model II | Model III | Model IV |
|---|---|---|---|---|---|
| $c_{11}$ | R | -0.0120*** (-7.6547) | -0.0121*** (-7.4074) | -0.0118*** (-7.3043) | -0.0112*** (-7.5434) |
| | G | -0.0121*** (-7.6443) | -0.0121*** (-6.8438) | -0.0122*** (-4.8472) | -0.0113*** (-6.8261) |
| $c_{12}$ | R | 2.71e-04 (1.4859) | 4.48 e-04* (1.8313) | 3.21 e-04 (1.0025) | 2.75 e -04 (1.2861) |
| | G | 2.46 e -04 (1.3581) | 4.46 e-04* (1.8074) | 3.64 e-04* (1.0732) | 2.95 e -04 (1.3597) |
| $c_{22}$ | R | 8.41e-04** (2.2754) | -5.94 e-04 (-1.6341) | 8.34 e-04 (1.5543) | -5.05 e-04 (-1.5533) |
| | G | 8.71e-04** (2.4581) | -5.69 e-04 (-1.4379) | 8.14 e-04 (1.4955) | 4.90 e-04 (-1.4334) |
| $a_{11}$ | R | 0.2626*** (6.8906) | 0.2321*** (5.8893) | 0.2414*** (6.0994) | 0.2367*** (6.5010) |
| | G | 0.2647*** (6.9755) | 0.2272*** (5.3746) | 0.2314*** (4.0545) | 0.2284*** (5.8924) |
| $a_{22}$ | R | 0.4598*** (13.3956) | 0.4669*** (15.3871) | 0.4523*** (13.8794) | 0.4758*** (13.8889) |
| | G | 0.4589*** (13.4889) | 0.4660*** (15.2226) | 0.4505*** (14.0148) | 0.4812*** (13.7421) |
| $b_{11}$ | R | 0.9229*** (96.477) | 0.9286*** (106.902) | 0.9292*** (110.283) | 0.9341*** (124.506) |
| | G | 0.9213*** (91.970) | 0.9296*** (106.559) | 0.9277*** (89.741) | 0.9352*** (125.463) |
| $b_{22}$ | R | 0.9006*** (87.067) | 0.9092*** (95.571) | 0.9180*** (94.528) | 0.9067*** (82.0706) |
| | G | 0.9064*** (88.189) | 0.9096*** (95.767) | 0.9180*** (95.379) | 0.9048*** (79.7415) |

Table 3
Mean equation estimates for state-dependent multi-factor models

The table shows the estimates of the risk-return trade-off using the state-dependent multi-factor model in Equation 5. Two versions of the model are estimated: R refers to the restricted version of the model in Equation 5 where $\lambda_{21,s_t} = \lambda_{22,s_t} = 0$ (similar to Scruggs, 1998) and G to the General model where all parameters are freely estimated. The proxy for the excess market returns ($r_{m,t}$) is the CRSP Value-Weighted portfolio minus the risk-free rate. Each column illustrates the results from the estimation of the model using one of the choices for the ($r_{b,t}$) alternative investment set. In model I the excess returns of the 5-year US Government Bond is used. In model II the excess returns of the 10-year US Government Bond is used. In model III the excess returns of the 20-year US Government Bond is used. In model IV the excess returns of an equally weighted portfolio containing these three bonds is used. The coefficients and corresponding t-statistics (in parenthesis) are shown for the full sample period (March 1953-June 2013). The t-statistics are computed using Bollerslev-Wooldridge standard errors ( *** , ** and * represents statistical significance at 1%, 5% and 10% levels). Panel A shows the estimations for the state $s_t=1$ which correspond to low volatility periods and Panel B shows the estimations for the state $s_t=2$ which correspond to high volatility periods.

$$r_{m,t,s_t} = \lambda_{10,s_t} + \lambda_{11,s_t}\sigma^2_{m,t,s_t} + \lambda_{12,s_t}\sigma_{mb,t,s_t} + \varepsilon_{m,t,s_t}$$

$$r_{b,t,s_t} = \lambda_{20,s_t} + \lambda_{21,s_t}\sigma_{bm,t,s_t} + \lambda_{22,s_t}\sigma^2_{b,t,s_t} + \varepsilon_{b,t,s_t}$$

| | | Model I | Model II | Model III | Model IV |
|---|---|---|---|---|---|
| colspan="6" | Panel A. Low volatility state ($s_t=1$) |
| $\lambda_{10,s_t=1}$ | R | 0.0103 (1.5267) | -0.0231* (-1.6875) | 0.0344*** (3.3647) | -0.0114 (-0.8301) |
| | G | 0.0040 (1.1357) | 0.0022 (0.4796) | 0.0022 (0.4709) | 0.0056 (1.3789) |
| $\lambda_{11,s_t=1}$ | R | 9.5670*** (2.7671) | 15.9695*** (4.7872) | 18.4561* (1.8612) | 5.3076** (2.3019) |
| | G | 9.7538*** (2.7423) | 12.1115*** (2.1571) | 12.1225** (1.9830) | 11.0457** (2.4170) |
| $\lambda_{12,s_t=1}$ | R | 5.0367 (0.7751) | 4.3367*** (4.6199) | 12.2170*** (2.8454) | 6.6429** (2.3237) |
| | G | 0.4954*** (2.4734) | -0.8795 (-1.1786) | -0.8765 (-1.0870) | -0.4749 (-0.3788) |
| $\lambda_{20,s_t=1}$ | R | 0.0011*** (11.2717) | 0.0017* (1.7776) | 0.0015 (1.1309) | 0.0021*** (2.6354) |
| | G | -0.0006 (-1.4039) | -0.0003 (0.3757) | -0.0003 (0.4019) | -0.0007 (-0.7458) |
| $\lambda_{21,s_t=1}$ | G | 0.2131*** (2.4831) | 0.2994 (0.5783) | 0.3006 (0.6380) | 1.7573 (0.7743) |
| $\lambda_{22,s_t=1}$ | G | 0.1442*** (3.0205) | 0.0663* (1.7225) | 0.0664* (1.8384) | 0.0922** (2.0885) |
| colspan="6" | Panel B. High volatility state ($s_t=2$) |
| $\lambda_{10,s_t=2}$ | R | 0.0036** (2.1084) | 0.0048 (1.3098) | 0.0024 (0.6351) | 0.0029 (0.7069) |
| | G | -0.0349* (-1.8797) | 0.0119 (1.3429) | 0.0191 (1.2923) | 0.0093 (1.1634) |
| $\lambda_{11,s_t=2}$ | R | 3.5663 (0.2264) | -9.6434* (1.6910) | -14.5513** (-2.1125) | -2.1357** (-2.1588) |
| | G | 5.8157 (0.8648) | -1.5298 (-0.1531) | -1.5737 (-0.1829) | -1.0780** (-1.9709) |
| $\lambda_{12,s_t=2}$ | R | -1.4201 (-0.9573) | -2.6444*** (-5.5163) | -9.1955*** (-2.7510) | 2.8581** (-2.3479) |
| | G | -0.4033 (-1.2127) | 2.9370 (1.3179) | 2.9276* (1.6938) | 2.9347 (0.6806) |
| $\lambda_{20,s_t=2}$ | R | -0.0004*** (-3.1656) | -0.0005 (-1.0752) | -0.0011** (-1.9850) | -0.0006 (-1.5629) |
| | G | -0.0014 (-0.6178) | -0.0003 (-0.1677) | -0.0004 (-0.1537) | 0.0013 (0.6275) |
| $\lambda_{21,s_t=2}$ | G | -0.4776** (-2.5848) | -0.3581 (-0.7942) | -0.3587 (-0.7379) | -1.1286 (-0.6595) |
| $\lambda_{22,s_t=2}$ | G | 0.0733 (0.7167) | -0.0039 (-0.1157) | 0.0039 (-0.1138) | 0.0007 (0.0183) |

Table 4
Variance equation estimates for state-dependent multi-factor models

The table shows the estimates of the variance equation of the state-dependent multi-factor model in Equation 6. These variance equations come from two different versions of the main model presented in equation 5: R refers to the restricted version of the model in Equation 5 where $\lambda_{21,st} = \lambda_{22,st} = 0$ (similar to Scruggs, 1998) and G to the General model where all parameters are freely estimated. The proxy for the excess market returns ($r_{m,t}$) is the CRSP Value-Weighted portfolio minus the risk-free rate. Each column illustrates the results from the estimation of the model using one of the choices for the ($r_{b,t}$) alternative investment set. In model I the excess returns of the 5-year US Government Bond is used. In model II the excess returns of the 10-year US Government Bond is used. In model III the excess returns of the 20-year US Government Bond is used. In model IV the excess returns of an equally weighted portfolio containing these three bonds is used. Within each alternative investment we distinguish between state $s_t=1$ which correspond to low volatility periods and state $s_t=2$ which correspond to high volatility periods. Excess monthly returns are used in the estimation of the model. The coefficients and corresponding t-statistics (in parenthesis) are shown for the full sample period (March 1953-June 2013). The t-statistics are computed using Bollerslev-Wooldridge standard errors ( $^{***}$, $^{**}$ and $^{*}$ represents statistical significance at 1%, 5% and 10% levels). The last two rows of the table shows the estimates for the parameters of the transition probability matrix in equation 7.

$$H_{t,s_t} = \begin{pmatrix} \sigma^2_{m,t,s_t} & \sigma_{mb,t,s_t} \\ \sigma_{mb,t,s_t} & \sigma^2_{b,t,s_t} \end{pmatrix} = \begin{pmatrix} c_{11,st} & 0 \\ c_{12,st} & c_{22,st} \end{pmatrix} \begin{pmatrix} c_{11,st} & 0 \\ c_{12,st} & c_{22,st} \end{pmatrix}' + \begin{pmatrix} a_{11,st} & a_{21,st} \\ a_{12,st} & a_{22,st} \end{pmatrix} \varepsilon_{t-1}\varepsilon_{t-1}' \begin{pmatrix} a_{11,st} & a_{21,st} \\ a_{12,st} & a_{22,st} \end{pmatrix}' + \begin{pmatrix} b_{11,st} & b_{21,st} \\ b_{12,st} & b_{22,st} \end{pmatrix} H_{t-1} \begin{pmatrix} b_{11,st} & b_{21,st} \\ b_{12,st} & b_{22,st} \end{pmatrix}' \quad for \ s_t = 1,2$$

| | | Model I | | Model II | | Model III | | Model IV | |
|---|---|---|---|---|---|---|---|---|---|
| | | $s_t = 1$ | $s_t = 2$ | $s_t = 1$ | $s_t = 2$ | $s_t = 1$ | $s_t = 2$ | $s_t = 1$ | $s_t = 2$ |
| $c_{11}$ | R | -0,0010$^{***}$ (-61.630) | 1,7228$^{***}$ (8,9911) | -0,5371$^{***}$ (-8.7636) | 2,2911$^{***}$ (12,232) | 0.0025 (0.0860) | 0.6885$^{***}$ (3.4195) | 0,0031 (0.2803) | 2,2174$^{***}$ (7.3845) |
| | G | 0.0011 (0.5530) | 3.2867$^{***}$ (6.9591) | 0.0001 (0.0396) | 1.6902$^{***}$ (3.9268) | -0.0001 (-0.0131) | 1.6900$^{***}$ (3.5581) | 0.0002 (0.0620) | 1.5366$^{***}$ (4.7093) |
| $c_{12}$ | R | 0,0000 (0.0066) | -0,0680$^{***}$ -(5,8563) | 0,0195 (0.0006) | -0,0831$^{***}$ (-11,248) | -0,0003 (-0.2108) | -0,0973$^{***}$ (-2.7354) | 0,0000 (-0.0006) | -0,0174$^{***}$ (-7.8907) |
| | G | 0.0000 (-0.0054) | -0.0540$^{*}$ (-1.8817) | 0.0000 (0.0001) | -0.0643 (0.0359) | 0.0000 (0.0001) | -0.0642$^{*}$ (1.8560) | 0.0000 (0.0010) | -0.0056 (0.0218) |
| $c_{22}$ | R | -0,0001 (-0.1161) | 0,2606$^{***}$ (5,6229) | -0,0319$^{**}$ (-2.2661) | 0,1361$^{*}$ (1,7949) | -0,0009 (-0.4217) | -0,2516$^{***}$ (-3.1059) | 0,0002 (0.1123) | 0,1204$^{**}$ (2.0302) |
| | G | 0.0002 (0.1434) | 0.4987$^{***}$ (4.6756) | 0.0000 (0.0172) | 0.1186 (1.5048) | -0.0000 (0.0058) | 0.1186 (1.4597) | 0.0001 (0.0256) | -0.1102$^{*}$ (1.6514) |
| $a_{11}$ | R | 0,0492$^{*}$ (1.9031)) | 0,0129 (0,9750) | 0,0002$^{***}$ (33.5008) | 0,0000$^{*}$ (1,9346) | 0.0060 (1.5856) | 0.0053 (1.4952) | 0,0001 (1.4762) | 0,0001$^{**}$ (2.1763) |
| | G | 0.1879$^{***}$ (3.7328) | 0.2216$^{***}$ (4.0342) | 0.0554 (1.3020) | 0.0082 (0.5007) | 0.0555$^{*}$ (1.6871) | 0.0081 (0.6502) | 0.0428 (1.1991) | 0.0180 (0.7784) |
| $a_{22}$ | R | 0,5786$^{***}$ (49.7512) | 0,1519$^{***}$ (25,4871) | 0,5367$^{***}$ (70.6927) | 0,0018$^{***}$ (4,0823) | 0.3485$^{***}$ (5.9286) | 0.3088$^{***}$ (5.5804) | 0,2128$^{*}$ (1.8658) | 0,1886$^{***}$ (2.7504) |
| | G | 0.3894$^{***}$ (0.0696) | 0.4548$^{***}$ (0.0753) | 0.5380 (1.4783) | 0.0796 (0.5685) | 0.5380$^{*}$ (1.8418) | 0.0799 (0.7100) | 0.5191$^{***}$ (4.9893) | 0.2187$^{***}$ (3.2385) |
| $b_{11}$ | R | 0,8845$^{***}$ (38.7541) | 1,1407$^{***}$ (41,8484) | 0,8224$^{***}$ (39.3749) | 1,0607$^{***}$ (44,7155) | 0.8241$^{***}$ (27.4831) | 1.2325$^{***}$ (33.6105) | 0,7048$^{***}$ (28.791) | 1,0534$^{***}$ (37.856) |
| | G | 0.8694$^{***}$ (14.4548) | 0.9312$^{***}$ (14.9593) | 0.8114$^{***}$ (17.4458) | 1.1237$^{***}$ (20.5306) | 0.8114$^{***}$ (16.9546) | 1.1237$^{***}$ (19.9531) | 0.7997$^{***}$ (20.9706) | 1.1499$^{***}$ (25.1493) |
| $b_{22}$ | R | 0,8318$^{***}$ (80.9501) | 1,0728$^{***}$ (87,6328) | 0,8245$^{***}$ (39.9644) | 1,0633$^{***}$ (45,3837) | 0.7878$^{***}$ (25.9462) | 1.1782$^{***}$ (31.7309) | 0,7077$^{***}$ (29.608) | 1,0576$^{***}$ (38.930) |
| | G | 0.8744$^{***}$ (15.4474) | 0.9365$^{***}$ (15.9567) | 0.7887$^{***}$ (23.7954) | 1.0922$^{***}$ (28.0029) | 0.7887$^{***}$ (23.2738) | 1.0923$^{***}$ (27.3899) | 0.7689$^{***}$ (25.4467) | 1.1055$^{***}$ (30.5161) |
| p | R | 0.8058$^{***}$ (31.7236) | | 0.8631$^{***}$ (35.3541) | | 0.7637$^{***}$ (29.3734) | | 0.8397$^{***}$ (23.0373) | |
| | G | 0.9176$^{***}$ (45.4736) | | 0.8072$^{***}$ (16.7065) | | 0.8061$^{***}$ (15.3766) | | 0.7892$^{***}$ (18.8192) | |
| q | R | 0.7258$^{***}$ (14.6885) | | 0.7820$^{***}$ (18.0196) | | 0.5998$^{***}$ (12.2302) | | 0.7860$^{***}$ (16.4037) | |
| | G | 0.7392$^{***}$ (12.976) | | 0.7444$^{***}$ (11.5441) | | 0.7443$^{***}$ (11.8855) | | 0.7310$^{***}$ (15.0042) | |

Table 5.
Mean equation estimates for SP500 using all multi-factor models

Panel A of this table shows the estimates of the risk-return trade-off using the single-regime multi-factor model in Equation 2 for the SP500 index as an alternative for the market portfolio. Panel B shows the estimates of the risk-return trade-off using the regime-switching multi-factor model in Equation 5 for the SP500 index as an alternative for the market portfolio. We show the results when we use the equally-weighted bond portfolio as the alternative investment set ($r_{b,t,st}$). Two versions of the model are estimated: R refers to the Restricted version of the model in Equations 2 and 5 where $\lambda_{21,st} = \lambda_{22,st} = 0$ (similar to Scruggs, 1998) and G to the General model where all parameters are freely estimated. Excess monthly returns are used in the estimation of the model. The coefficients and corresponding t-statistics (in parenthesis) are shown for the full sample period (March 1953-June 2013). The t-statistics are computed using Bollerslev-Wooldridge standard errors (***, ** and * represents statistical significance at 1%, 5% and 10% levels).

| | | | $r_{m,t,s_t} = \lambda_{10,s_t} + \lambda_{11,s_t}\sigma^2_{m,t,s_t} + \lambda_{12,s_t}\sigma_{mb,t,s_t} + \varepsilon_{m,t,s_t}$ $r_{b,t,s_t} = \lambda_{20,s_t} + \lambda_{21,s_t}\sigma_{bm,t,s_t} + \lambda_{22,s_t}\sigma^2_{b,t,s_t} + \varepsilon_{b,t,s_t}$ | | | | | |
|---|---|---|---|---|---|---|---|
| | | Panel A. Single-Regime model | Panel B. Regime-Switching model | | | | |
| | | Single-Regime | Low volatility state ($s_t = 1$) | | | High volatility state ($s_t = 2$) | |
| $\lambda_{10}$ | R | -0.0091 (-1.1900) | $\lambda_{10,s_t=1}$ | R | -0.0053 (-0.1212) | $\lambda_{10,s_t=2}$ | R | -0.0133 (-0.8295) |
| | G | -0.0130 (1.4140) | | G | 0.0022 (0.4911) | | G | 0.0119 (1.3024) |
| $\lambda_{11}$ | R | 6.3818 (1.5054) | $\lambda_{11,s_t=1}$ | R | 18.3502*** (7.0532) | $\lambda_{11,s_t=2}$ | R | -14.1840*** (-10.5751) |
| | G | 8.7912* (1.7107) | | G | 12.0803** (2.1870) | | G | -1.3256 (-0.1340) |
| $\lambda_{12}$ | R | 0.3822 (0.0434) | $\lambda_{12,s_t=1}$ | R | -4.5477*** (-9.3330) | $\lambda_{12,s_t=2}$ | R | 5.8620*** (6.4722) |
| | G | 2.2695 (0.2478) | | G | -0.8841 (-0.8525) | | G | 2.9801 (1.1734) |
| $\lambda_{20}$ | R | 0.0003 (0.7771) | $\lambda_{20,s_t=1}$ | R | -0.0519 (-1.2481) | $\lambda_{20,s_t=2}$ | R | 0.0025** (2.2982) |
| | G | 0.0001 (0.2144) | | G | -0.0003 (-0.4379) | | G | -0.0003 (-0.1524) |
| $\lambda_{21}$ | G | 0.8167* (1.8533) | $\lambda_{21,s_t=1}$ | G | 0.2910 (0.6721) | $\lambda_{21,s_t=2}$ | G | -0.3516 (-0.8434) |
| $\lambda_{22}$ | G | 0.4861** (2.5831) | $\lambda_{22,s_t=1}$ | G | 0.0660** (2.0519) | $\lambda_{22,s_t=2}$ | G | -0.0040 (-0.1220) |

Table 6.
Mean equation estimates for multi-factor models (single regime and state-dependent) using the databases in Scruggs (1998) and Scruggs and Glabadanidis (2003)

The table shows the estimates of the risk-return trade-off using the independent multi-factor model in Equation 2 (panel A) and the state-dependent multi-factor model in Equation 5 (panels B and C). Two versions of these model are estimated: R refers to the Restricted version of the model in Equations 2 and 5 where $\lambda_{21,(st)} = \lambda_{22,(st)} = 0$ (Scruggs, 1998) and G refers to the General model where all parameters are freely estimated. Excess monthly returns are used in the estimation of the model. The proxy for the excess market returns ($r_{m,t}$) in Scruggs (1998) is the CRSP Value-Weighted portfolio of NYSE-AMEX stocks minus the 1-month T-bill yield from March 1950 to December 1994 while in Scruggs and Glabadanidis (2003) the sample period is from January 1953 to December 1997. The choice for the ($r_{b,t}$) alternative investment is the Ibboston Associates long-term U.S. Treasury bond total return index for the corresponding sample periods. The coefficients and corresponding t-statistics (in parenthesis) are shown for the full sample period (March 1950 to December 1994 in Scruggs (1998) and January 1953 to December 1997 in Scruggs and Glabadanidis (2003)). The t-statistics are computed using Bollerslev-Wooldridge standard errors (***, ** and * represents statistical significance at 1%, 5% and 10% levels).

$$r_{m,t} = \lambda_{10} + \lambda_{11}\sigma^2_{m,t} + \lambda_{12}\sigma_{mb,t} + \varepsilon_{m,t}$$
$$r_{b,t} = \lambda_{20} + \lambda_{21}\sigma_{bm,t} + \lambda_{22}\sigma^2_{b,t} + \varepsilon_{b,t}$$

**Panel A. Linear model**

| | | $\lambda_{10}$ | $\lambda_{11}$ | $\lambda_{12}$ | $\lambda_{20}$ | $\lambda_{21}$ | $\lambda_{22}$ |
|---|---|---|---|---|---|---|---|
| Scruggs (1998) | R | -0.0031 (-0.4818) | 5.3664 (1.3292) | -2.3777 (-0.5064) | -0.0005 (-0.1166) | - | - |
| | G | -0.0039 (-0.6214) | 6.3774 (1.6151) | -0.8979 (-0.1936) | -0.0004 (-0.8502) | 3.4492 (0.8792) | -0.0102 (1.4094) |
| Scruggs and Glabad (2003) | R | 0.0049 (1.0071) | 3.0546 (0.9958) | 1.2023 (0.2661) | 0.0002 (0.4674) | - | - |
| | G | -0.0010 (-0.1968) | 4.7449 (1.4200) | -0.2603 (-0.0583) | -0.0001 (-0.2862) | 5.2473 (1.4286) | 1.7210 (0.9746) |

$$r_{m,t,s_t} = \lambda_{10,s_t} + \lambda_{11,s_t}\sigma^2_{m,t,s_t} + \lambda_{12,s_t}\sigma_{mb,t,s_t} + \varepsilon_{m,t,s_t}$$
$$r_{b,t,s_t} = \lambda_{20,s_t} + \lambda_{21,s_t}\sigma_{bm,t,s_t} + \lambda_{22,s_t}\sigma^2_{b,t,s_t} + \varepsilon_{b,t,s_t}$$

**Panel B. Low volatility state ($s_t=1$)**

| | | $\lambda_{10,s_t=2}$ | $\lambda_{11,s_t=2}$ | $\lambda_{12,s_t=2}$ | $\lambda_{20,s_t=2}$ | $\lambda_{21,s_t=2}$ | $\lambda_{22,s_t=2}$ |
|---|---|---|---|---|---|---|---|
| Scruggs (1998) | R | 0.0042 (1.0890) | 12.9126*** (2.8804) | -0.4506*** (-2.8024) | -0.0004 (-0.7623) | - | - |
| | G | -0.0037 (-0.8085) | 20.6704*** (3.5105) | -0.4743** (-2.6360) | 0.0004 (1.1624) | 0.2251* (1.7996) | -0.0200 (-0.6630) |
| Scruggs and Glabad (2003) | R | 0,0064 (1.7984) | 10.7221*** (2.2399) | -0.4187*** (-2.6746) | -0.0001 (-0.2364) | - | - |
| | G | 0.0055* (1.8425) | 17.1605*** (3.4377) | -0.5750** (-2.5650) | -0.0002 (-0.3057) | 0.4077** (1.9683) | -0.0565 (-1.1828) |

$$r_{m,t,s_t} = \lambda_{10,s_t} + \lambda_{11,s_t}\sigma^2_{m,t,s_t} + \lambda_{12,s_t}\sigma_{mb,t,s_t} + \varepsilon_{m,t,s_t}$$
$$r_{b,t,s_t} = \lambda_{20,s_t} + \lambda_{21,s_t}\sigma_{bm,t,s_t} + \lambda_{22,s_t}\sigma^2_{b,t,s_t} + \varepsilon_{b,t,s_t}$$

**Panel C. High volatility state ($s_t=2$)**

| | | $\lambda_{10,s_t=2}$ | $\lambda_{11,s_t=2}$ | $\lambda_{12,s_t=2}$ | $\lambda_{20,s_t=2}$ | $\lambda_{21,s_t=2}$ | $\lambda_{22,s_t=2}$ |
|---|---|---|---|---|---|---|---|
| Scruggs (1998) | R | 0.0103 (1.3912) | -14.1498*** (-2.7193) | 0.5594*** (2.8069) | 0.0002 (0.1602) | - | - |
| | G | 0.0029*** (2.8503) | -26.8595*** (-3.6500) | 0.6760*** (2.9173) | -0.0005*** (-3.9145) | -0.3050** (-1.9867) | 0.1134** (2.3156) |
| Scruggs and Glabad (2003) | R | 0.0110 (1.4134) | -15.0876** (-2.5637) | 0.5437*** (2.7185) | 0.0003 (0.2498) | - | - |
| | G | 0.0149* (1.9581) | -18.5588*** (-2.7977) | 0.6116** (2.6324) | -0.0008 (-0.4956) | -0.1738 (-1.2269) | 0.0572 (1.2168) |

Table 7.
Mean equation estimates for linear models controlling for high-volatility observations
The table shows the estimates of the risk-return trade-off using the single-regime multi-factor model in Equation 2 if we control for the high volatility observations with a dummy variable $D_i$ (taking value 1 for high volatility states) for each series using equations 8-11. Two versions of the model are estimated: R refers to the restricted version of the model in Equation 2 where $\lambda_{21} = \lambda_{22} = 0$ (Scruggs, 1998) and G to the General model where all parameters are freely estimated. Excess monthly returns are used in the estimation of the model and the sample periods follows previous analysis. We estimate the model for several alternatives of the market portfolio. Each column illustrates the results from the estimation of the model using one of these alternatives as the excess market returns ($r_{m,t}$): CRSP Value-Weighted portfolio, the SP500 index and the datasets used in Scruggs (1998) and Scruggs and Glabadanidis (2003). In the first 2 columns we show the results when we use the equally-weighted bond portfolio as the alternative investment set ($r_{b,t}$). In the other 2 columns the alternative investment set is the one used in the corresponding paper. The t-statistics are computed using Bollerslev-Wooldridge standard errors (***, ** and * represents statistical significance at 1%, 5% and 10% levels).

$$r_{m,t} = \lambda_{10} + \lambda_{11}\sigma^2_{m,t} + \lambda_{11d}D_1\sigma^2_{m,t} + \lambda_{12}\sigma_{mb,t} + \lambda_{12d}D_2\sigma_{mb,t} + \varepsilon_{m,t}$$

$$r_{b,t} = \lambda_{20} + \lambda_{21}\sigma_{bm,t} + \lambda_{22}\sigma^2_{b,t} + \varepsilon_{b,t}$$

| | | CRSP | SP500 | Scruggs | Scruggs and Glab |
|---|---|---|---|---|---|
| $\lambda_{10}$ | R | -0.0142 (-1.5622) | -0.0437*** (-4.1424) | -0.2499 (-0.3679) | 0.4716 (0.9311) |
| | G | -0.0152* (-1.7542) | -0.0334*** (-5.3779) | -0.2949 (-0.4250) | 0.3439 (0.6459) |
| $\lambda_{11}$ | R | 23.1338*** (3.3699) | 41.9584*** (5.1381) | 13.1768** (2.4876) | 7.7647** (1.9737) |
| | G | 25.1016*** (3.7785) | 34.6001*** (9.2975) | 13.6446** (2.4966) | 9.1683** (2.2090) |
| $\lambda_{11,d}$ | R | -25.4618*** (-8.6418) | -44.2197*** (-12.1802) | -33.6692*** (-7.2032) | -32.7354*** (-7.6828) |
| | G | -27.0210*** (-9.5708) | -43.9572*** (-12.7881) | -33.7009*** (-7.3179) | -32.8019*** (-7.4893) |
| $\lambda_{12}$ | R | -25.6111 (-1.6164) | 6.6709 (0.3813) | -23.8281*** (-3.3738) | -22.8791*** (-3.4201) |
| | G | 3.2608 (0.2320) | 6.4367 (0.4881) | -20.8688*** (-20.9354) | -20.1689*** (-2.9303) |
| $\lambda_{12,d}$ | R | 53.4563** (2.4028) | 3.5437 (0.1075) | 71.0805*** (4.3543) | 74.3805*** (5.1215) |
| | G | 18.8511 (0.9687) | 7.7326 (0.2849) | 70.1197*** (4.3739) | 74.5994*** (4.8167) |
| $\lambda_{20}$ | R | 0.0002 (0.4852) | 0.0001 (0.2954) | 0.0009 (0.0192) | 0.0137 (0.2770) |
| | G | 0.0004 (0.8718) | 0.0004 (0.0868) | 0.0472 (0.9272) | 0.0280 (0.5018) |
| $\lambda_{21}$ | G | 2.0084 (0.3132) | 12.4749* (1.8751) | 0.4549 (0.1074) | 5.4795 (1.2761) |
| $\lambda_{22}$ | G | 5.4200** (2.3794) | 4.7549** (2.5954) | 3.0327* (1.7532) | 1.6108 (0.8732) |

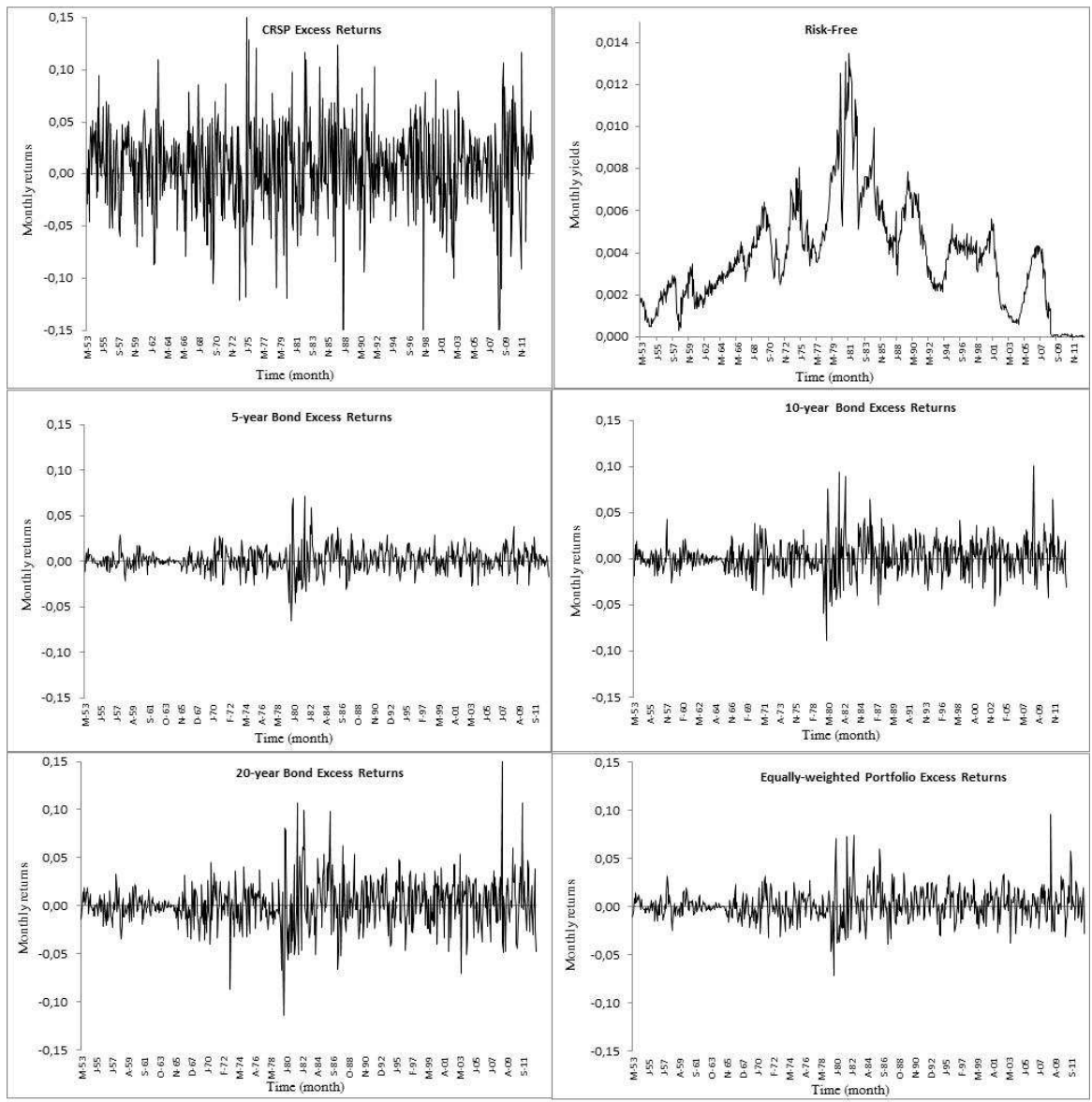

Figure 1. Monthly excess returns of the selected portfolios
The figure plots the excess monthly returns on the series selected for the market portfolio ($r_{m,t}$) and the alternative investment sets ($r_{b,t}$). The proxy for the market portfolio is the CRSP value-weighted NYSE-AMEX portfolio and the risk-free asset is the yield on the one-month Treasury bill. The proxies for the alternative opportunity set are the total returns (computed as Morningstar ®) for the 5,10 and 20 years constant maturity US Government Bonds and an equally weighted portfolio containing these three bonds. The sample period includes observations from 1953 to 2013 and the monthly returns and yields are expressed as decimals.

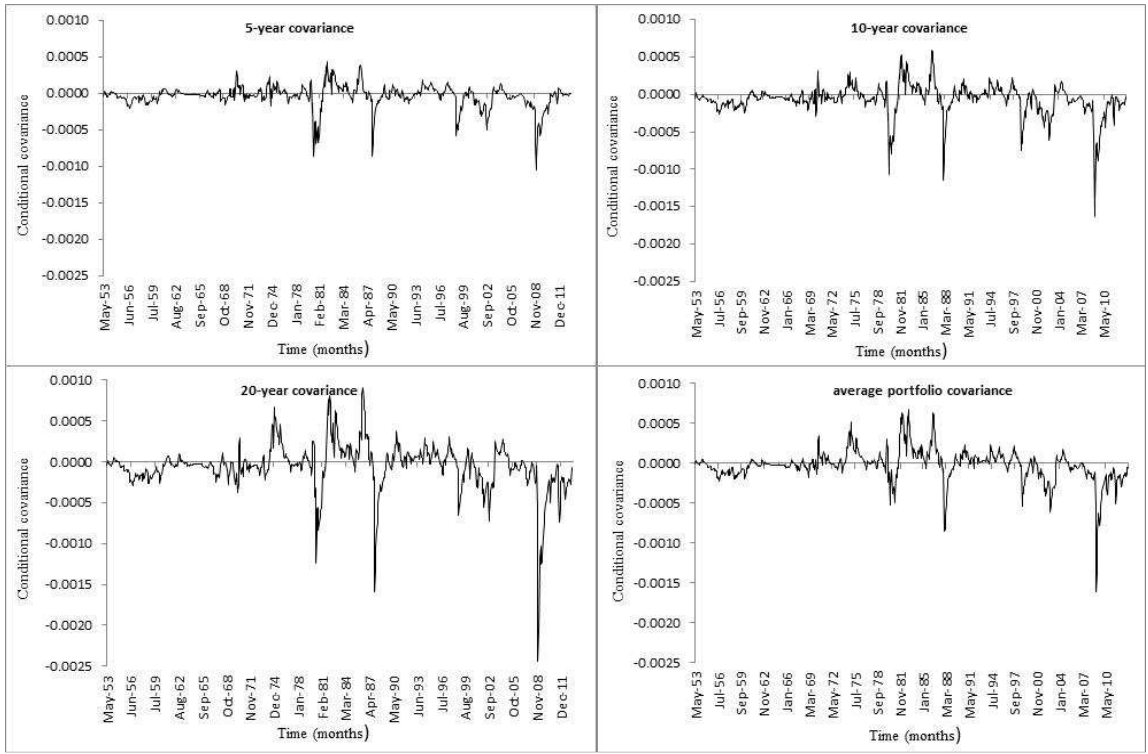

Figure 2. Conditional covariances between excess market returns and the intertemporal component
The figure plots the conditional covariances ($\sigma_{mb,t}$) from the multifactor model in Equation 6. In these plots we use the CRSP value-weighted NYSE-AMEX portfolio for the excess market returns ($r_{m,t}$) and the total returns (computed as Morningstar ®) for the 5,10, 20 years constant maturity US Government Bonds and an equally weighted portfolio containing these three bonds as alternative investment sets ($r_{b,t}$). The sample period includes observations from 1953 to 2013 and the monthly conditional covariances are expressed as decimals.

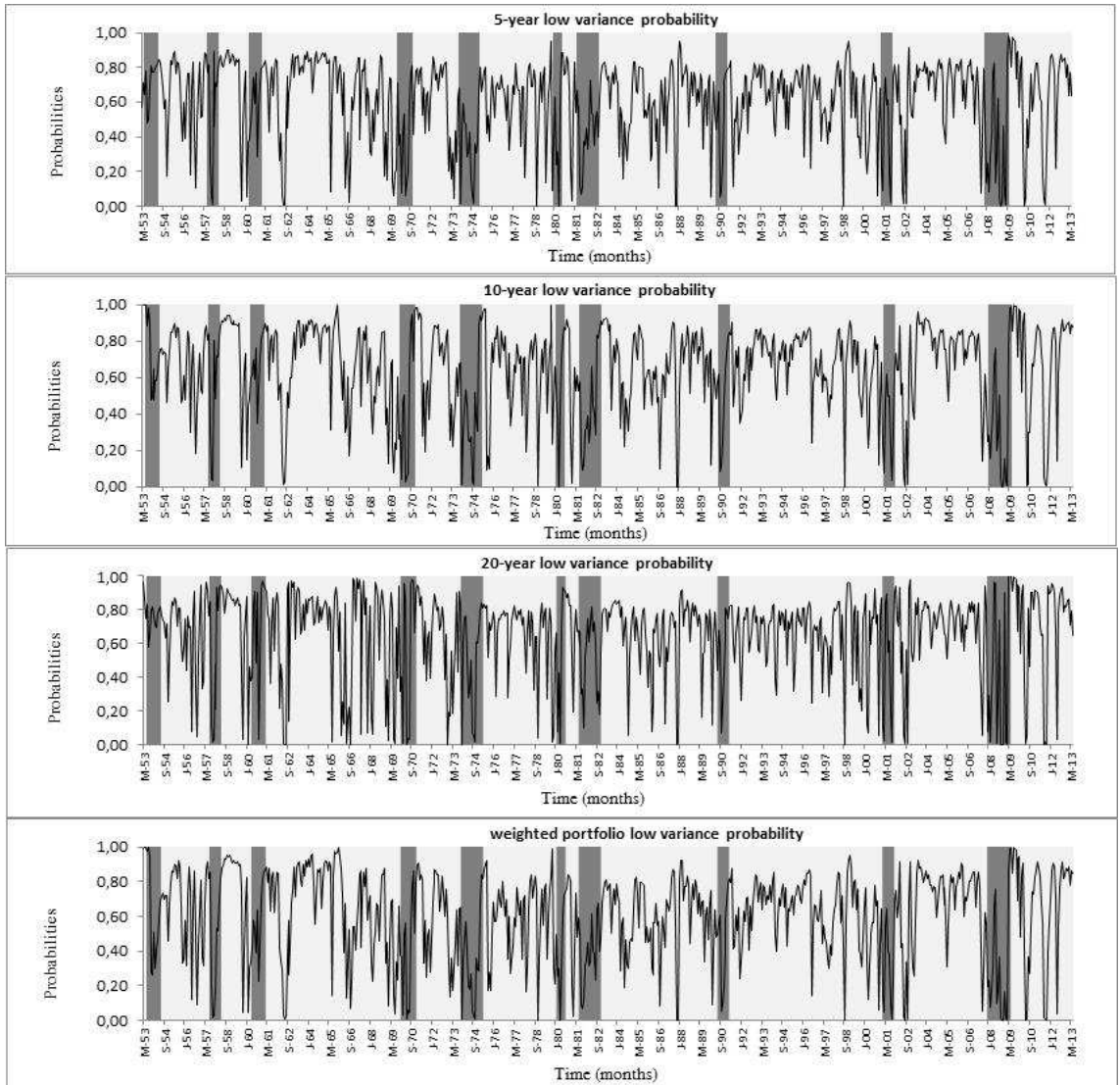

Figure 3. Filtered probabilities for low volatility states
This figure plots the probability of being in a low probability state [$P(s_t=1/\Omega_t,\theta)$] for the state-dependent multifactor model according to Equation (10). In these plots we use the CRSP value-weighted NYSE-AMEX portfolio for the excess market returns ($r_{m,t}$) and the total returns (computed as Morningstar ®) for the 5,10, 20 years constant maturity US Government Bonds and an equally weighted portfolio containing the three previous bonds as alternative investment sets ($r_{b,t}$). Shaded areas correspond to NBER recessionary periods. The sample period includes observations from 1953 to 2013.

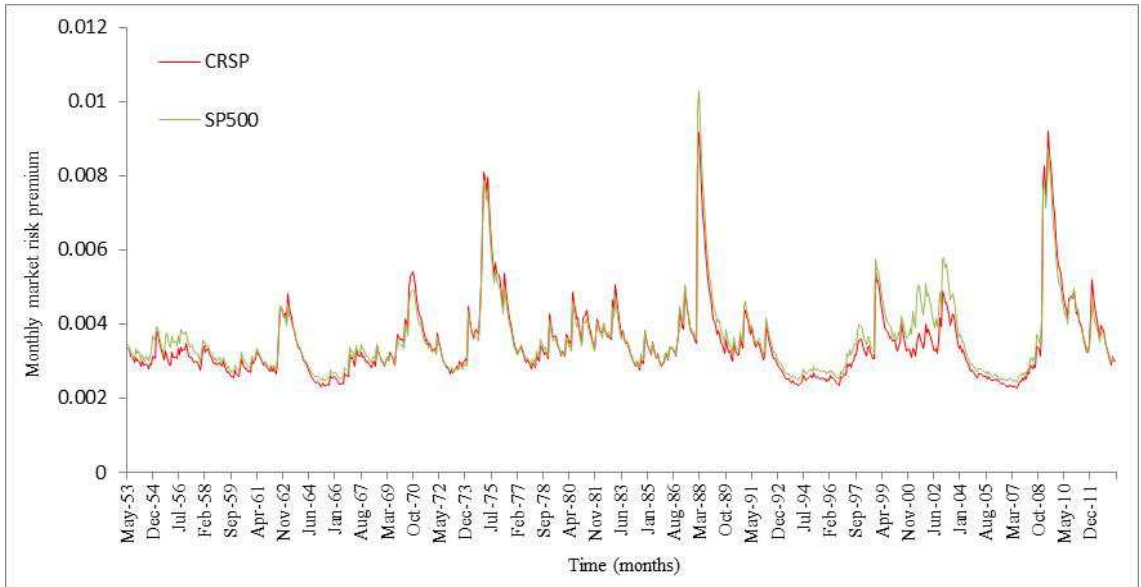

Figure 4.A. Market risk premium for the linear multi-factor model
This figure plots the estimated risk premium for different portfolios using the linear multi-factor model in equation 2. The green line represents SP500 and the red line is the premium associated with the CRSP Value Weighted Index Portfolio. The sample period covers observations from 1953 to 2013 and the monthly risk premiums are expressed as decimals.

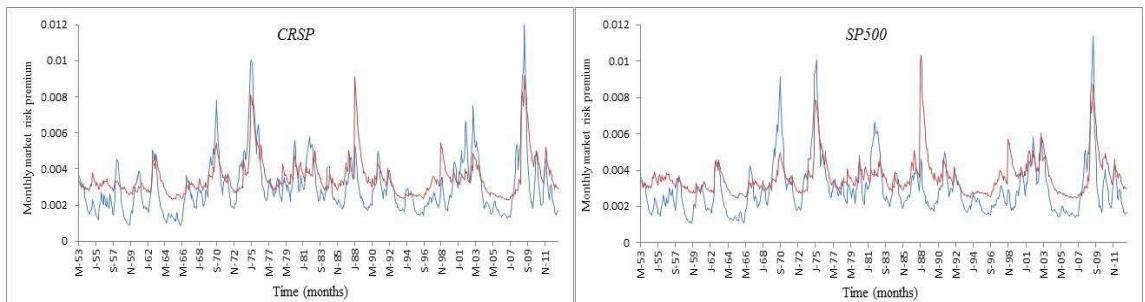

Figure 4.B Market risk premium comparison between linear and non-linear models
This figure plots for each market portfolio (CRSP and SP500) the estimated risk premiums using the linear multi-factor model in equation 2 (red line) and using the state-dependent multi-factor model in equation 5 (blue line). The sample period covers observations from 1953 to 2013 and the monthly risk premiums are expressed as decimals.

31